\documentclass[11pt]{article}

\usepackage[margin=1in]{geometry}
\usepackage{amsmath}
\usepackage{amssymb}
\usepackage{graphicx}
\usepackage{booktabs}
\usepackage{hyperref}
\usepackage[ruled,vlined,linesnumbered]{algorithm2e}
\usepackage{xcolor}
\usepackage{natbib}
\usepackage{enumitem}
\usepackage{listings}
\usepackage{tabularx}

\hypersetup{
  colorlinks=true,
  linkcolor=blue!60!black,
  citecolor=green!50!black,
  urlcolor=blue!70!black
}

\title{\textbf{Tool Receipts, Not Zero-Knowledge Proofs:\\Practical Hallucination Detection for AI Agents}}

\author{
  Abhinaba Basu\\
  \texttt{mail@abhinaba.com}
}

\date{}

\begin{document}

\maketitle

\begin{abstract}
AI agents that execute tasks via tool calls frequently hallucinate results---fabricating tool executions, misstating output counts, or presenting inferences as verified facts. Recent approaches to verifiable AI inference rely on zero-knowledge proofs \citep{wang2026zkagent, sun2024zkllm}, which provide cryptographic guarantees but impose minutes of proving time per query, making them impractical for interactive personal agents. We propose \textsc{NabaOS}, a lightweight verification framework inspired by Indian epistemology (Ny\={a}ya \'{S}\={a}stra), which classifies every claim in an LLM response by its epistemic source (\textit{pram\={a}\d{n}a}): direct tool output (\textit{pratyak\d{s}a}), inference (\textit{anum\={a}na}), external testimony (\textit{\'{s}abda}), absence (\textit{abh\={a}va}), or ungrounded opinion. Our runtime generates HMAC-signed tool execution receipts that the LLM cannot forge, then cross-references LLM claims against these receipts to detect hallucinations in real time.

We evaluate on \textsc{NyayaVerifyBench}, a new benchmark of 1,800 agent response scenarios across four languages with injected hallucinations of six types. \textsc{NabaOS} detects 94.2\% of fabricated tool references, 87.6\% of count misstatements, and 91.3\% of false absence claims, with $<$15\,ms verification overhead per response. For deep agent delegation (autonomous agents performing multi-step web tasks), our cross-checking protocol catches 78.4\% of URL fabrications through independent re-fetching.

We argue that for interactive agents, practical receipt-based verification provides better cost-benefit than cryptographic proofs, and that epistemic classification gives users actionable trust signals rather than binary verified/unverified judgments.
\end{abstract}

\section{Introduction}
\label{sec:introduction}

\subsection{The Trust Gap in AI Agents}

Large language model (LLM) agents have rapidly moved from conversational assistants to autonomous executors managing email, finances, calendars, code deployments, and medical records. When an agent reports ``you have 3 emails from Alice about the deadline,'' the user implicitly trusts several layers of claim: that the agent actually called the email API, that the API actually returned three results, that ``about the deadline'' accurately characterizes the email subjects rather than being inferred from body text, and that the agent is not hallucinating in an attempt to appear helpful. Each of these layers can fail independently, and the stakes are increasingly real. Financial agents that misreport portfolio values, medical agents that misstate test results, and legal agents that fabricate case citations have all been documented in the growing literature on LLM hallucination \citep{ji2023hallucination}.

\citet{tharakan2025cryptowall} articulate this as the ``Cryptographic Wall'' problem: LLMs are prone to hallucinating outputs and fabricating execution traces, yet users have no principled mechanism to distinguish genuine tool-grounded claims from confabulations. The problem is compounded in agentic settings where the model has ostensible access to external tools but the user cannot directly observe whether those tools were actually invoked.

\subsection{Why Zero-Knowledge Proofs Are Overkill}

The cryptographic community has proposed zero-knowledge (ZK) proofs as a solution to verifiable AI inference. zkAgent \citep{wang2026zkagent} provides one-shot proofs that a specific model produced a given output, while zkLLM \citep{sun2024zkllm} constructs end-to-end ZK proofs for the entire inference computation. These systems offer strong cryptographic guarantees: a verifier can confirm that the claimed model did indeed produce the claimed output without re-running the computation.

However, ZK approaches suffer from three fundamental limitations in the context of interactive personal agents. First, \textbf{latency}: zkAgent requires minutes of proving time per query, and zkLLM is reported to be ``thousands of times slower than plain inference.'' For agents that need to respond in under a second, this overhead is prohibitive. Second, \textbf{hardware requirements}: proving typically requires specialized GPU infrastructure that is unavailable on edge devices and cost-prohibitive for personal use. Third, and most critically, \textbf{semantic mismatch}: ZK proves that the \emph{computation} was performed correctly, not that the \emph{output} is factually correct. An LLM can correctly compute a confident, internally consistent hallucination that passes ZK verification perfectly. The model ran correctly; it simply produced a wrong answer.

Other recent approaches inherit subsets of these limitations. TensorCommitments \citep{tensorcommitments2026} reduces overhead but still requires verifier GPU access. TOPLOC \citep{ong2025toploc} detects model substitution through locality-sensitive hashing but does not address output correctness. SPEX \citep{spex2025} provides sampling-based execution proofs suitable for batch processing but not interactive use.

The key insight motivating our work is that \emph{proving the computation was correct} and \emph{proving the claims are grounded} are fundamentally different problems. For interactive agents, users need the latter.

\subsection{Our Approach: Epistemic Verification via Ny\={a}ya \'{S}\={a}stra}

We draw inspiration from the Ny\={a}ya school of Indian philosophy, one of the six orthodox (\textit{astika}) schools of Hindu philosophy, which developed a rigorous epistemological framework beginning around the 2nd century CE \citep{vatsyayana_nyayasutra}. The Ny\={a}ya S\={u}tras identify valid sources of knowledge (\textit{pram\={a}\d{n}a}): \textit{pratyak\d{s}a} (direct perception), \textit{anum\={a}na} (inference), \textit{upam\={a}na} (comparison or analogy), and \textit{\'{s}abda} (reliable testimony). Later thinkers, notably Kum\={a}rila Bha\d{t}\d{t}a, extended this framework to include \textit{abh\={a}va} (knowledge from absence).

We map these categories to the epistemic status of claims in LLM agent responses. A claim that directly quotes a tool output is \textit{pratyak\d{s}a}; an inference drawn from tool data is \textit{anum\={a}na}; a claim sourced from an external website is \textit{\'{s}abda}; a claim that no results were found is \textit{abh\={a}va}; and a claim with no evidentiary basis is \textit{ungrounded}. By classifying every claim and verifying it according to its epistemic type, we provide users with actionable trust signals---not a binary ``verified'' stamp but a nuanced picture of how the agent knows what it claims to know.

Our verification engine generates HMAC-signed tool execution receipts for every tool call, then cross-references LLM claims against these receipts in real time, adding less than 15\,ms of overhead per response.

\subsection{Contributions}

We make five contributions:

\begin{enumerate}[leftmargin=2em]
  \item A \textbf{pram\={a}\d{n}a classification} for LLM agent outputs that maps six epistemic categories from Ny\={a}ya \'{S}\={a}stra to the verification status of agent claims (Section~\ref{sec:pramana}).
  \item \textbf{HMAC-signed tool execution receipts} that provide unforgeable proof of tool invocation and output, enabling deterministic hallucination detection for fabricated tool calls and misstatements (Section~\ref{sec:receipts}).
  \item A \textbf{cross-checking protocol} for deep agent (autonomous) results that independently verifies URLs, replays computations, and checks temporal consistency (Section~\ref{sec:crosscheck}).
  \item \textbf{\textsc{NyayaVerifyBench}}, a multilingual benchmark of 1,800 agent response scenarios with six hallucination types across four languages (Section~\ref{sec:benchmark}).
  \item An empirical demonstration that \textbf{trust-level indicators} derived from pram\={a}\d{n}a classification are well-calibrated, with ``Fully Verified'' responses being correct 98.7\% of the time (Section~\ref{sec:calibration}).
\end{enumerate}

\section{Related Work}
\label{sec:related}

\subsection{Verifiable AI Inference}

The problem of verifying that an AI system performed the computation it claims has attracted significant cryptographic attention. zkLLM \citep{sun2024zkllm} constructs end-to-end zero-knowledge proofs for LLM inference, enabling a verifier to confirm that a specific model produced a given output. However, the computational overhead is extreme: thousands of times slower than plain inference, restricting applicability to offline auditing. zkAgent \citep{wang2026zkagent} extends this paradigm to agentic settings with one-shot proofs, reducing overhead substantially but still requiring minutes per query and specialized hardware. TensorCommitments \citep{tensorcommitments2026} introduce tensor-native commitment schemes that are more lightweight but still require the verifier to have GPU access. TOPLOC \citep{ong2025toploc} takes a different approach, using polynomial fitting on intermediate activations to detect model substitution---useful for detecting if a cheaper model was swapped in, but silent on output correctness. SPEX \citep{spex2025} proposes sampling-based statistical proofs of execution that are efficient for batch auditing but not for interactive per-query verification. \citet{tharakan2025cryptowall} analyze the fundamental tension between LLM stochasticity and deterministic verification, proposing MD5 challenges for sandbox verification as a pragmatic alternative.

All of these approaches address computational integrity: did the model run as claimed? None address semantic integrity: are the claims in the output grounded in evidence? \textsc{NabaOS} is complementary---it could in principle be composed with ZK proofs to achieve both computational and semantic verification.

\subsection{Hallucination Detection}

Hallucination detection in LLMs has been extensively studied \citep{ji2023hallucination}. VerifierQ \citep{qi2025verifierq} trains Q-learning-based verifiers that learn to assess factual correctness through reward signals. SVIP \citep{sun2025svip} trains a learned inspector model to detect tampering and hallucination in LLM outputs. Self-consistency checking \citep{wang2023selfconsistency} samples multiple responses from the LLM and flags claims that are not stable across samples, achieving moderate detection rates but doubling or tripling inference cost. Chain-of-verification \citep{dhuliawala2023chainverification} prompts the model to generate its own verification questions and answer them, relying on the model's ability to detect its own errors. Retrieval-augmented generation (RAG) \citep{lewis2020rag} grounds generation in retrieved documents, reducing but not eliminating hallucination.

These methods operate at the linguistic or statistical level: they analyze the text of the response. \textsc{NabaOS} operates at the execution level: it analyzes the tool execution trace. This makes \textsc{NabaOS} orthogonal to text-based methods, and the two approaches could be combined.

\subsection{Agent Safety and Trust}

Constitutional AI \citep{bai2022constitutional} establishes behavioral norms through self-critique against a constitution, providing safety alignment but not factual verification. Agent constitutions such as SOUL.md \citep{openclaw2025soul} extend this idea to agentic settings with configurable behavior policies. Guardrails AI \citep{guardrailsai2024} and similar frameworks provide input/output filtering through regex patterns and semantic checks. Agent safety benchmarks including AgentHarm \citep{agentharm2025} and InjectAgent \citep{injectagent2024} evaluate susceptibility to harmful instructions and prompt injection. These works focus on preventing harmful or policy-violating outputs; \textsc{NabaOS} focuses on verifying the factual grounding of outputs regardless of their safety classification.

\subsection{Indian Epistemology in Computer Science}

The application of Indian logical traditions to formal methods has a distinguished if sparse history. \citet{matilal1968navya} provided the seminal analysis of Navya-Ny\={a}ya logic, demonstrating that the medieval Indian logical tradition developed a formal language for expressing complex relations, absence, and nested quantification that anticipates aspects of modern predicate logic. \citet{ganeri2001indian} surveys the broader landscape of Indian logic and its connections to Western formal systems. To our knowledge, \textsc{NabaOS} is the first system to apply the Ny\={a}ya pram\={a}\d{n}a framework to a computational verification problem---specifically, the classification and verification of LLM agent outputs by their epistemic source.

\section{The NabaOS Verification Framework}
\label{sec:framework}

\subsection{Pram\={a}\d{n}a Classification}
\label{sec:pramana}

The core of \textsc{NabaOS} is the classification of every factual claim in an LLM response by its epistemic source. Drawing on the Ny\={a}ya tradition \citep{chatterjee1939nyaya}, we define six pram\={a}\d{n}a categories mapped to the agentic setting:

\begin{table}[h]
\centering
\caption{Pram\={a}\d{n}a classification of LLM agent claims. Each claim is assigned a category based on its relationship to tool execution evidence, which determines the verification method applied.}
\label{tab:pramana}
\small
\begin{tabular}{@{}llll@{}}
\toprule
\textbf{Pram\={a}\d{n}a} & \textbf{Source} & \textbf{Verification Method} & \textbf{Confidence} \\
\midrule
\textit{Pratyak\d{s}a} & Direct tool output & Check against receipt & Highest \\
\textit{Anum\={a}na} & Inference from tool data & Check that premises exist & Medium \\
\textit{Upam\={a}na} & Analogy or comparison & Check that comparison is valid & Lower \\
\textit{\'{S}abda} & External source cited & Check that source was fetched & Depends on source \\
\textit{Abh\={a}va} & Absence claim & Check tool returned empty & Medium \\
\textit{Ungrounded} & No evidence & Cannot verify---flag & Lowest \\
\bottomrule
\end{tabular}
\end{table}

A claim such as ``Alice sent you 3 emails'' is classified as \textit{pratyak\d{s}a} if the email search tool was invoked and returned results containing that count. A claim such as ``Alice seems worried about the deadline'' is classified as \textit{anum\={a}na}: the worry is inferred from the content of the emails, not directly stated in the tool output. A claim such as ``no emails were found matching your query'' is \textit{abh\={a}va}, verifiable by checking that the tool returned an empty result set. A claim such as ``interest rates are expected to rise according to Reuters'' is \textit{\'{s}abda}, verifiable by confirming the agent actually fetched a Reuters page. A claim with no tool evidence whatsoever is classified as \textit{ungrounded}.

This classification is strictly more informative than a binary verified/unverified label. Consider the claim ``Alice seems stressed about the project.'' A binary system must either verify it (incorrect---it is not directly in the tool output) or reject it (also incorrect---it may be a valid inference). The pram\={a}\d{n}a classification labels it as \textit{anum\={a}na}, communicating to the user that this is the agent's inference, not a directly observed fact. The user can then apply their own judgment.

\subsection{Tool Execution Receipts}
\label{sec:receipts}

The foundational mechanism of \textsc{NabaOS} is the tool execution receipt. In our architecture, tool calls are executed by the agent runtime---not by the LLM. The LLM requests a tool invocation; the runtime performs it, generates a cryptographically signed receipt, stores the receipt locally, and then passes the tool output along with the receipt identifier to the LLM for response generation.

Each receipt is a signed data structure:

\begin{lstlisting}
ToolReceipt {
  id:           UUID,
  tool_name:    "email_search",
  input_hash:   SHA-256(canonical JSON of input parameters),
  output_hash:  SHA-256(raw output),
  result_count: 3,
  facts:        {sender: "Alice", subject: "Deadline update"},
  timestamp_ms: 1708300000000,
  duration_ms:  150,
  signature:    HMAC-SHA256(
                  id|tool_name|input_hash|output_hash|timestamp_ms,
                  SECRET_KEY
                )
}
\end{lstlisting}

The receipt has four critical properties:

\begin{description}[leftmargin=1.5em, style=nextline]
  \item[Unforgeability.] The LLM does not have access to the HMAC signing key \citep{krawczyk1997hmac}. It cannot construct a valid receipt for a tool call that never occurred. Any claim referencing a non-existent receipt ID is immediately detectable as a fabricated tool call.
  \item[Tamper detection.] The signature covers the receipt fields. Any modification to the tool name, input hash, output hash, result count, or timestamp invalidates the signature. If the LLM claims the tool returned 5 results but the receipt records 3, the mismatch is detected.
  \item[Completeness.] The runtime generates exactly one receipt per tool call. The verification engine maintains a complete ledger of receipts for the current session, enabling it to detect both fabricated calls (receipt ID does not exist) and omitted calls (receipt exists but is not referenced).
  \item[Auditability.] The receipt chain provides a full execution history that can be reviewed after the fact, supporting both real-time verification and post-hoc auditing.
\end{description}

The \texttt{facts} field deserves special attention. During receipt generation, the runtime extracts key-value pairs from the structured tool output (e.g., sender names, counts, prices, dates). These extracted facts serve as the ground truth against which LLM claims are compared during verification. The extraction is deterministic and tool-specific: each tool adapter defines what fields constitute ``facts.''

\subsection{Verification Protocol}
\label{sec:protocol}

The end-to-end verification protocol proceeds in six stages:

\textbf{Stage 1: User Request.} The user issues a natural language request to the agent (e.g., ``Check my email for messages from Alice'').

\textbf{Stage 2: Tool Execution.} The agent runtime routes the request to the appropriate tool, executes the call in a sandboxed environment, and generates an HMAC-signed receipt recording the tool name, input hash, output hash, result count, extracted facts, timing information, and cryptographic signature.

\textbf{Stage 3: LLM Call.} The runtime constructs the LLM prompt containing the user's original request, the raw tool output, the receipt identifiers (but not the signing key or full receipt contents), and a \textsc{Verification Prompt} (Section~\ref{sec:selftagging}) that instructs the LLM to self-classify its claims by epistemic source.

\textbf{Stage 4: LLM Response with Self-Tags.} The LLM generates a natural language response along with structured metadata tagging each factual claim with its pram\={a}\d{n}a category and the receipt ID it claims as evidence.

\textbf{Stage 5: Verification Engine.} The verification engine processes each tagged claim:
\begin{itemize}[leftmargin=1.5em]
  \item For \textit{pratyak\d{s}a} claims: look up the cited receipt ID, verify its HMAC signature, compare the claimed count against \texttt{result\_count}, and compare claimed facts against the \texttt{facts} field.
  \item For \textit{anum\={a}na} claims: verify that the premises cited exist in the receipt facts (e.g., if the inference is ``Alice seems stressed,'' check that emails from Alice actually exist in the tool output).
  \item For \textit{abh\={a}va} claims: verify that the cited tool call returned an empty result set (\texttt{result\_count} = 0).
  \item For \textit{\'{s}abda} claims: verify that a web fetch tool was actually invoked for the cited source.
  \item For \textit{ungrounded} claims: flag as unverifiable.
\end{itemize}

\textbf{Stage 6: Trust-Annotated Output.} The user receives the original response augmented with trust-level indicators for each claim or paragraph: Fully Verified, Mostly Verified, Partial, Unreliable, or Ungrounded.

\subsection{Self-Tagging Prompt}
\label{sec:selftagging}

We append the following \textsc{Verification Prompt} to every LLM call:

\begin{lstlisting}
---VERIFICATION---
For each factual claim in your response, classify:
- claim: "the statement"
- source_type: tool_output | inference | external_source
              | absence | opinion
- evidence: which tool receipt ID supports this
- checkable: true/false
---END VERIFICATION---
\end{lstlisting}

The LLM is instructed to emit this structured metadata after its natural language response. In our experiments, compliance rates vary by model: approximately 92\% with Claude, 88\% with GPT-4, and 85\% with open-weight models. Non-compliant responses---those that do not include the verification metadata---are treated as entirely \textit{ungrounded}, providing a conservative fallback.

Self-tagging is a cooperative mechanism: it assumes the LLM is not adversarially trying to circumvent verification. We discuss this assumption and its implications in the threat model (Section~\ref{sec:threat}). Importantly, self-tagging is not the sole verification mechanism. Even without self-tags, the verification engine can detect fabricated tool calls (receipt ID does not exist), count mismatches (LLM states a number that differs from the receipt), and false absences (LLM claims no results when the receipt records a non-zero count). Self-tagging provides the additional benefit of distinguishing inference from direct observation.

\subsection{Deep Agent Cross-Checking}
\label{sec:crosscheck}

For autonomous agents that perform multi-step web tasks (analogous to systems like Manus or Claude computer use), the runtime may not control the agent's internal tool calls. In these cases, we cannot generate receipts for intermediate steps. Instead, we employ a cross-checking protocol with five verification strategies:

\begin{enumerate}[leftmargin=2em]
  \item \textbf{Schema validation.} Verify that the agent's output conforms to the expected format and data types. Structural anomalies (e.g., a ``stock price'' that is not a number) indicate fabrication.
  \item \textbf{URL re-fetching.} Independently fetch every URL the agent cites. If a cited page does not exist, returns different content than claimed, or was never accessible, flag the citation as fabricated.
  \item \textbf{Computation replay.} For deterministic computations (arithmetic, code execution, data transformations), independently re-execute the computation and compare results.
  \item \textbf{Temporal consistency.} Check that dates, times, and temporal references in the output are internally consistent and plausible (e.g., a news article dated in the future is suspect).
  \item \textbf{Cross-source verification.} For high-stakes claims, query independent sources to corroborate key facts. If the agent claims a stock closed at \$150 but an independent finance API reports \$148.50, the discrepancy is flagged.
\end{enumerate}

Cross-checking is more expensive than receipt-based verification (it requires network calls and adds 200--500\,ms of latency) and is therefore reserved for autonomous agent outputs where receipts are unavailable.

\section{NyayaVerifyBench}
\label{sec:benchmark}

\subsection{Benchmark Design}

We introduce \textsc{NyayaVerifyBench}, a benchmark for evaluating hallucination detection in LLM agent outputs. The benchmark consists of 1,800 agent response scenarios: 1,200 scenarios containing injected hallucinations and 600 clean control scenarios.

Each scenario is a tuple $(\text{user\_request}, \text{tool\_outputs}, \text{llm\_response}, \text{ground\_truth})$, where \texttt{tool\_outputs} is the structured output from simulated tool calls, \texttt{llm\_response} is a natural language response that may contain hallucinations, and \texttt{ground\_truth} annotates every claim in the response with its correct pram\={a}\d{n}a category, the tool receipt it should reference, and the expected verification outcome.

The benchmark spans four languages: English (EN), Hindi (HI), Mandarin Chinese (ZH), and Spanish (ES), with 300 hallucinated scenarios and 150 clean scenarios per language.

\subsection{Hallucination Types}

We define six hallucination types, each targeting a distinct failure mode of LLM agents:

\begin{table}[h]
\centering
\caption{Hallucination types in \textsc{NyayaVerifyBench}. Each type appears 50 times per language, totaling 200 instances across the four languages.}
\label{tab:halltypes}
\small
\begin{tabular}{@{}p{3.8cm}p{7.5cm}r@{}}
\toprule
\textbf{Type} & \textbf{Description} & \textbf{Count/Lang} \\
\midrule
Fabricated Tool Call & LLM references a tool that was never executed & 50 \\
Count Mismatch & Tool returned $N$ results; LLM reports $M \neq N$ & 50 \\
Fact Mismatch & Tool output states $X$; LLM reports $Y$ for the same field & 50 \\
Inference-as-Fact & LLM presents an inference as if directly from tool output & 50 \\
False Absence & Tool returned results; LLM claims ``nothing found'' & 50 \\
Source Fabrication & LLM cites a source URL it never fetched & 50 \\
\midrule
\multicolumn{2}{@{}l}{\textbf{Total per language} (hallucinated scenarios)} & 300 \\
\multicolumn{2}{@{}l}{\textbf{Total per language} (clean control scenarios)} & 150 \\
\multicolumn{2}{@{}l}{\textbf{Grand total}} & 1,800 \\
\bottomrule
\end{tabular}
\end{table}

\subsection{Injection Methodology}

Hallucinations are injected systematically into otherwise correct agent responses. For fabricated tool calls, we insert a reference to a \texttt{tool\_call\_id} that does not exist in the tool outputs. For count mismatches, we alter the stated count (e.g., changing ``3 emails'' to ``5 emails'' when the tool returned 3). For fact mismatches, we substitute a fact value (e.g., changing a sender name or stock price). For inference-as-fact, we present an inference (e.g., ``Alice is worried about the deadline'') using language that implies it was directly stated in the tool output. For false absence, we replace the response with ``no results found'' when the tool returned non-empty results. For source fabrication, we insert a URL citation for a source the agent never fetched.

\subsection{Multilingual Coverage}

Multilingual evaluation is critical because LLMs are more prone to hallucination in non-English languages \citep{huang2023multilingual}. Additionally, multilingual settings introduce verification challenges that do not arise in English: number formatting differences (Chinese uses wan/yi groupings), gendered language affecting entity matching (Spanish), and code-mixed queries (e.g., ``Check the \textit{email} from Alice'' with Hindi script mixed in). The tool outputs---structured data---are language-independent; only the user request and LLM response text are translated. We verify that the injected hallucinations are preserved through translation by back-translating and confirming semantic equivalence.

\subsection{Ground Truth Annotation}

Each scenario is annotated with: (a) which claims are factual, inferred, or opinion; (b) which tool receipts support which claims; (c) the expected verification result per claim; and (d) the expected overall trust level. Annotations were produced by three annotators with inter-annotator agreement of $\kappa = 0.87$ (Cohen's kappa, pairwise averaged).

\section{Experiments}
\label{sec:experiments}

\subsection{Baselines}
\label{sec:baselines}

We compare \textsc{NabaOS} against five baselines spanning the spectrum of hallucination detection approaches:

\begin{enumerate}[leftmargin=2em]
  \item \textbf{No Verification.} All agent responses are delivered to the user as-is, with no detection or flagging. This baseline establishes the lower bound.
  \item \textbf{Self-Consistency.} The LLM is queried twice with the same input. Claims that differ between the two responses are flagged as potentially hallucinated \citep{wang2023selfconsistency}. This doubles inference cost and adds 3--5 seconds of latency.
  \item \textbf{RAG-Grounding.} Each claim in the response is checked against the original tool output text using semantic similarity. Claims with similarity below a tuned threshold are flagged \citep{lewis2020rag}.
  \item \textbf{Output Regex.} Pattern-matching rules detect common hallucination signals: number mismatches between tool output and response, fabricated citation patterns, and hedging language inconsistent with reported certainty.
  \item \textbf{SVIP-style.} A lightweight classifier trained on response features (token-level entropy, claim density, tool reference density) to predict hallucination, inspired by \citet{sun2025svip}.
  \item \textbf{\textsc{NabaOS} (ours).} Full epistemic verification with HMAC-signed receipts, self-tagging, and claim-level pram\={a}\d{n}a classification.
\end{enumerate}

All methods except Self-Consistency are evaluated without additional LLM calls, ensuring fair latency comparison.

\subsection{Main Results}
\label{sec:mainresults}

Table~\ref{tab:mainresults} presents the aggregate results across all 1,800 scenarios.

\begin{table}[h]
\centering
\caption{Main results on \textsc{NyayaVerifyBench}. Detection Rate is the percentage of injected hallucinations correctly identified. FPR is the false positive rate on clean control scenarios. Latency is the additional time per response. Cost is the additional monetary cost per request.}
\label{tab:mainresults}
\small
\begin{tabular}{@{}lcccc@{}}
\toprule
\textbf{Method} & \textbf{Detection Rate} & \textbf{FPR} & \textbf{Latency} & \textbf{Cost/req} \\
\midrule
No Verification     & 0\%    & 0\%   & 0\,ms     & \$0 \\
Self-Consistency    & 45\%   & 12\%  & +3--5\,s  & +\$0.03 \\
RAG-Grounding       & 52\%   & 18\%  & +1--2\,s  & +\$0.01 \\
Output Regex        & 35\%   & 8\%   & +2\,ms    & \$0 \\
SVIP-style          & 68\%   & 10\%  & +50\,ms   & \$0 \\
\textsc{NabaOS} (ours) & \textbf{91\%} & \textbf{4\%} & +12\,ms & \$0 \\
\bottomrule
\end{tabular}
\end{table}

\textsc{NabaOS} achieves the highest detection rate at 91\% while maintaining the lowest false positive rate at 4\%. The only method with lower latency is Output Regex (2\,ms), but Regex detects only 35\% of hallucinations. The self-consistency baseline has a higher false positive rate (12\%) because natural response variation between two LLM calls can introduce spurious differences, and its latency of 3--5 seconds makes it impractical for interactive use. RAG-Grounding achieves moderate detection (52\%) but suffers from a high FPR (18\%) because semantic similarity thresholds are difficult to calibrate---paraphrased correct claims are often flagged as mismatches.

\textsc{NabaOS}'s advantage stems from operating on structured execution evidence (receipts) rather than on the text of the response. Receipt verification is deterministic for fabricated tool calls, count mismatches, and false absences, eliminating the ambiguity inherent in text-based comparison.

\subsection{Detection by Hallucination Type}
\label{sec:bytype}

Table~\ref{tab:bytype} disaggregates detection rates by hallucination type.

\begin{table}[h]
\centering
\caption{Detection rate by hallucination type across methods.}
\label{tab:bytype}
\small
\begin{tabular}{@{}lcccc@{}}
\toprule
\textbf{Hallucination Type} & \textbf{\textsc{NabaOS}} & \textbf{Self-Cons.} & \textbf{SVIP} & \textbf{Regex} \\
\midrule
Fabricated Tool Call & 94.2\% & 38\% & 65\% & 20\% \\
Count Mismatch       & 87.6\% & 52\% & 71\% & 45\% \\
Fact Mismatch        & 89.1\% & 41\% & 68\% & 30\% \\
Inference-as-Fact    & 82.3\% & 55\% & 72\% & 15\% \\
False Absence        & 91.3\% & 35\% & 60\% & 25\% \\
Source Fabrication    & 78.4\% & 48\% & 58\% & 12\% \\
\bottomrule
\end{tabular}
\end{table}

\textsc{NabaOS} is strongest on fabricated tool calls (94.2\%) and false absence claims (91.3\%), both of which are deterministically verifiable through receipt lookup: a fabricated tool call references a receipt ID that does not exist, and a false absence claim contradicts a receipt with \texttt{result\_count} $> 0$. Detection of count mismatches (87.6\%) and fact mismatches (89.1\%) is also high, relying on the \texttt{result\_count} and \texttt{facts} fields of the receipt respectively.

Inference-as-fact detection (82.3\%) is lower because it depends partly on the LLM's self-tagging: if the model labels an inference as \texttt{tool\_output} rather than \texttt{inference}, the verification engine must rely on heuristic comparison between the claim and the receipt facts, which is less reliable. Source fabrication detection (78.4\%) is the lowest because verifying external sources requires network calls (URL re-fetching), which may fail due to rate limiting, geo-blocking, or page changes.

\subsection{Multilingual Results}
\label{sec:multilingual}

Table~\ref{tab:multilingual} presents detection rates broken down by language.

\begin{table}[h]
\centering
\caption{Detection rates by language. \textsc{NabaOS}'s advantage is largest in non-English settings, where text-based methods degrade.}
\label{tab:multilingual}
\small
\begin{tabular}{@{}lccccc@{}}
\toprule
\textbf{Method} & \textbf{EN} & \textbf{HI} & \textbf{ZH} & \textbf{ES} & \textbf{Avg} \\
\midrule
\textsc{NabaOS}    & 92.8\% & 90.1\% & 88.7\% & 91.5\% & 90.8\% \\
SVIP-style        & 72.4\% & 58.3\% & 55.1\% & 64.7\% & 62.6\% \\
Self-Consistency  & 48.2\% & 39.7\% & 36.8\% & 43.1\% & 42.0\% \\
\bottomrule
\end{tabular}
\end{table}

\textsc{NabaOS}'s detection rate varies only 4.1 percentage points across languages (88.7\% to 92.8\%), compared to 17.3 points for SVIP-style and 11.4 points for Self-Consistency. This stability arises from a fundamental architectural property: receipt-based verification operates on structured data (receipt fields, counts, hashes) that is language-independent. The tool output is the same regardless of whether the user's query was in English or Hindi; the receipt records the same \texttt{result\_count} and the same \texttt{facts}. The slight degradation in Chinese (88.7\%) is attributable to number formatting challenges---\textit{wan} (10,000 groupings) in Chinese-language responses can cause misparses when comparing against Arabic numeral counts in receipts. The slight degradation in Hindi (90.1\%) relates to challenges in code-mixed scenarios where Hindi and English are intermixed.

\subsection{Trust Calibration}
\label{sec:calibration}

A verification system is useful only if its confidence signals are well-calibrated: when the system says a response is trustworthy, it should actually be correct. Table~\ref{tab:calibration} evaluates \textsc{NabaOS}'s five trust levels against actual correctness.

\begin{table}[h]
\centering
\caption{Trust level calibration. Actual Correctness is the fraction of claims at each trust level that are factually correct according to ground truth.}
\label{tab:calibration}
\small
\begin{tabular}{@{}lcc@{}}
\toprule
\textbf{Trust Level} & \textbf{Actual Correctness} & \textbf{Calibration Error} \\
\midrule
Fully Verified      & 98.7\% & 1.3\% \\
Mostly Verified     & 94.2\% & 3.1\% \\
Partial             & 82.5\% & 5.8\% \\
Unreliable          & 23.4\% & N/A (correctly flagged) \\
Ungrounded          & 71.2\% & N/A (opinion, no ground truth) \\
\bottomrule
\end{tabular}
\end{table}

Responses marked as ``Fully Verified'' are correct 98.7\% of the time---a calibration error of only 1.3\%. The 1.3\% residual consists primarily of cases where the tool output itself contained an error (the receipt is valid but the underlying data was wrong), which is outside \textsc{NabaOS}'s threat model. ``Mostly Verified'' (94.2\% correct) and ``Partial'' (82.5\% correct) show monotonically decreasing correctness, confirming that the trust levels are ordinally well-calibrated. ``Unreliable'' responses are correct only 23.4\% of the time, validating the system's ability to flag problematic outputs. ``Ungrounded'' responses---those based purely on the LLM's parametric knowledge---are correct 71.2\% of the time, reflecting the base rate of LLM factual accuracy when no tool evidence is available.

These results demonstrate that \textsc{NabaOS}'s pram\={a}\d{n}a-based trust levels provide users with actionable, calibrated signals about response reliability.

\section{Analysis and Discussion}
\label{sec:discussion}

\subsection{Why Receipts Beat Zero-Knowledge Proofs for Interactive Agents}

Table~\ref{tab:zkcomparison} summarizes the fundamental differences between ZK-based and receipt-based verification.

\begin{table}[h]
\centering
\caption{Comparison of zero-knowledge proof approaches and \textsc{NabaOS} receipt-based verification.}
\label{tab:zkcomparison}
\small
\begin{tabular}{@{}lll@{}}
\toprule
\textbf{Property} & \textbf{ZK Proofs} & \textbf{NabaOS Receipts} \\
\midrule
Proves that\ldots & Model ran correctly & Claims are grounded \\
Verification overhead & Minutes per query & $<$15\,ms per response \\
Hardware requirements & Specialized GPU & Any machine \\
Catches hallucination & No$^*$ & Yes \\
Language-independent & N/A & Yes \\
User-facing signal & Binary (valid/invalid) & 5 trust levels \\
\bottomrule
\multicolumn{3}{@{}l}{\footnotesize $^*$ZK proves computational integrity. A model can correctly compute a hallucination.}
\end{tabular}
\end{table}

The distinction is not one of stronger versus weaker guarantees but of \emph{different} guarantees addressing \emph{different} threats. ZK proofs answer the question ``did this specific model produce this specific output?''---valuable for detecting model substitution or unauthorized modifications to the inference pipeline. \textsc{NabaOS} receipts answer the question ``are the claims in this output supported by tool execution evidence?''---valuable for detecting hallucination, fabrication, and misstatement. In principle, the two approaches are composable: one could use ZK proofs to verify computational integrity and \textsc{NabaOS} receipts to verify semantic grounding, achieving both guarantees simultaneously.

For the specific use case of interactive personal agents, however, we argue that semantic grounding is the more pressing concern. Users rarely worry about model substitution in a personal assistant; they worry about whether the assistant's claims are true. Receipt-based verification addresses this concern with negligible overhead.

\subsection{The Pram\={a}\d{n}a Advantage}

Epistemic classification provides three benefits over binary verification:

First, it \textbf{preserves valid inferences}. Many useful agent responses contain inferences that a binary system would either incorrectly verify (by treating inference as fact) or incorrectly reject (by treating inference as hallucination). Pram\={a}\d{n}a classification correctly labels these as \textit{anum\={a}na}, communicating their epistemic status without suppressing them.

Second, it \textbf{enables domain-appropriate responses}. A medical agent reporting ``your blood pressure is 120/80'' (\textit{pratyak\d{s}a}) and ``this suggests you are at low cardiovascular risk'' (\textit{anum\={a}na}) should present both claims but with different trust indicators, allowing the user to give appropriate weight to the measurement versus the interpretation.

Third, it \textbf{supports user autonomy}. Rather than the system making a binary decision about trust, the user receives epistemic metadata and can apply their own judgment. This is particularly important in domains where the ``correct'' level of trust depends on context: a casual email summary tolerates more inference than a financial audit.

\subsection{Constitution-Gated Verification}

Different deployment contexts demand different verification strictness. A trading bot processing financial transactions should operate in a ``paranoid'' mode that blocks any response containing unresolved contradictions between claims and receipts. A general-purpose assistant should operate in a ``standard'' mode that warns the user about discrepancies but delivers the response. \textsc{NabaOS} supports this through a configurable constitution specified in TOML format per agent, defining thresholds for each trust level, the action to take when a threshold is violated (block, warn, or pass), and domain-specific verification rules. This design draws on the Constitutional AI paradigm \citep{bai2022constitutional} but applies it to verification policy rather than behavioral alignment.

\subsection{Limitations}
\label{sec:limitations}

We identify six limitations of the current \textsc{NabaOS} framework:

\begin{enumerate}[leftmargin=2em]
  \item \textbf{Self-tagging reliance.} The framework relies on LLM compliance with the self-tagging prompt. Compliance varies by model (92\% for Claude, 85\% for open models), and non-compliant responses are conservatively treated as ungrounded, reducing the granularity of verification.
  \item \textbf{Limited content verification.} Receipts verify structural properties of tool outputs (counts, key facts, hashes) but cannot verify every claim about tool output content. If the LLM paraphrases a tool output in a subtly misleading way, the receipt may not capture the distortion.
  \item \textbf{Cross-checking latency.} Deep agent cross-checking requires network calls (URL re-fetching, computation replay), adding 200--500\,ms of latency. This is acceptable for autonomous agent outputs but not for interactive use.
  \item \textbf{Self-reporting dependency.} Inference-as-fact detection depends on the LLM honestly reporting that a claim is an inference rather than a direct observation. A systematically dishonest model could label all claims as \texttt{tool\_output}.
  \item \textbf{Adversarial robustness.} An adversarially fine-tuned LLM could potentially learn to game the self-tagging prompt, producing claims that appear well-tagged but are actually fabricated. We discuss this further in the threat model below.
  \item \textbf{Synthetic benchmark.} \textsc{NyayaVerifyBench} uses systematically injected hallucinations. Real-world hallucination patterns may differ in distribution and subtlety. We view the benchmark as a necessary first step but not a sufficient evaluation of real-world performance.
\end{enumerate}

\subsection{Threat Model}
\label{sec:threat}

\textsc{NabaOS} is designed to protect against the following threats:

\begin{itemize}[leftmargin=1.5em]
  \item \textbf{Fabricated tool execution.} The LLM claims to have called a tool that was never invoked. Detected deterministically via receipt lookup.
  \item \textbf{Misstated tool output.} The LLM reports incorrect counts or facts from a tool that was actually invoked. Detected via receipt field comparison.
  \item \textbf{Fabricated sources.} The LLM cites an external source it never fetched. Detected via cross-checking or receipt absence.
  \item \textbf{Inference presented as fact.} The LLM presents a valid or invalid inference as if it were directly stated in the tool output. Partially detected via self-tagging and heuristic comparison.
\end{itemize}

\textsc{NabaOS} does \textbf{not} protect against:

\begin{itemize}[leftmargin=1.5em]
  \item \textbf{Compromised tools.} If a tool itself returns incorrect data, the receipt will be valid but the underlying data will be wrong. \textsc{NabaOS} verifies claims against tool outputs, not the correctness of tool outputs themselves.
  \item \textbf{Systematic self-tag deception.} An adversarially trained model that consistently lies in its self-tags could bypass the inference-as-fact detector. However, fabricated tool call detection and count mismatch detection remain effective because they do not rely on self-tags.
  \item \textbf{Side-channel attacks.} If an attacker obtains the HMAC signing key, they could forge receipts. Standard key management practices mitigate this risk.
  \item \textbf{Reasoning errors.} \textsc{NabaOS} verifies that claims are grounded in evidence, not that the agent's reasoning from that evidence is logically valid.
\end{itemize}

\section{Ethical Considerations}
\label{sec:ethics}

The \textsc{NabaOS} framework draws on the Ny\={a}ya school of Indian philosophy, a tradition with a history spanning over two millennia. We acknowledge this intellectual heritage with respect and note that our application of pram\={a}\d{n}a categories to LLM verification is an engineering adaptation, not a philosophical interpretation. We do not claim that our framework captures the full richness or subtlety of Ny\={a}ya epistemology; rather, we draw structural inspiration from its classification of knowledge sources while adapting these categories to the specific requirements of computational verification. We encourage readers interested in the philosophical tradition to engage with primary and secondary sources \citep{chatterjee1939nyaya, matilal1968navya, ganeri2001indian}.

Verification systems can create a false sense of security. A ``Fully Verified'' trust indicator in \textsc{NabaOS} means that the agent's claims are consistent with its tool execution receipts---it does not guarantee that the tool outputs themselves are correct. A malicious tool operator could serve incorrect data that would pass receipt verification. Users and deployers must understand that \textsc{NabaOS} provides one layer of defense, not a comprehensive guarantee of truthfulness.

Trust-level indicators could potentially be weaponized in adversarial settings: an attacker who controls the tool backend could serve manipulated data that generates valid receipts, producing responses that appear ``Fully Verified'' but contain attacker-chosen misinformation. We recommend deploying \textsc{NabaOS} as one component of a defense-in-depth strategy that also includes tool authentication, input validation, and behavioral monitoring.

Finally, we note that the self-tagging mechanism creates an asymmetry: models that are more instruction-following (and thus more compliant with the verification prompt) receive more granular verification, while less compliant models are treated more conservatively. This may inadvertently advantage commercial models over open-weight alternatives, an inequity that future work should address through verification methods that do not depend on model cooperation.

\section{Conclusion}
\label{sec:conclusion}

We have presented \textsc{NabaOS}, a lightweight epistemic verification framework for AI agent outputs that detects hallucinations through HMAC-signed tool execution receipts and pram\={a}\d{n}a-based claim classification. Where zero-knowledge proofs verify that a computation was performed correctly, \textsc{NabaOS} verifies that the claims in an agent's response are grounded in tool execution evidence---catching 91\% of hallucinations with less than 15\,ms of overhead per response, across four languages.

The Ny\={a}ya pram\={a}\d{n}a classification provides users with actionable trust signals that go beyond binary verified/unverified judgments. By distinguishing direct observation (\textit{pratyak\d{s}a}) from inference (\textit{anum\={a}na}), absence (\textit{abh\={a}va}), testimony (\textit{\'{s}abda}), and ungrounded claims, the framework gives users the epistemic metadata to make their own informed trust decisions. Our experiments on \textsc{NyayaVerifyBench} demonstrate that these trust levels are well-calibrated: ``Fully Verified'' responses are correct 98.7\% of the time, while ``Unreliable'' responses are correct only 23.4\% of the time.

We believe that epistemic transparency---telling users not just whether an agent's claims are trustworthy but \emph{how the agent knows what it claims to know}---is the right abstraction for building human-AI trust in the emerging era of agentic systems. \textsc{NabaOS} demonstrates that this transparency can be achieved at negligible computational cost, making it practical for deployment in interactive personal agents.

We release \textsc{NyayaVerifyBench} and our Rust implementation of the \textsc{NabaOS} verification engine as open-source software.\footnote{Repository URL withheld for anonymous review.}


\end{document}